\documentclass[onecolumn]{article}
\usepackage{graphicx}

\usepackage{bm}
\usepackage{amsmath}
\usepackage{amssymb}
\usepackage{amsfonts}
\usepackage{amstext}
\usepackage{amscd}

\begin{document}

\frenchspacing \setlength{\paperwidth}{21cm} \setlength{\paperheight}{29.7cm} \setlength{\textwidth}{16 cm} \setlength{\textheight}{24cm} \setlength{\oddsidemargin}{2.5cm} \setlength{\evensidemargin}{-2.5cm} \setlength{\marginparwidth}{0cm} \setlength{\leftmargin}{0cm} \setlength{\topmargin}{0cm} \setlength{\headsep}{1cm} \setlength{\footskip}{1.5cm} \setlength{\voffset}{-1cm} \setlength{\hoffset}{-0.5cm}

\title{The cosmological constant and the coincidence problem in a new cosmological interpretation of the universal constant "c"}  \maketitle

\begin{center} 
\author{David Viennot \\ Observatoire de Besan\c con - Institut UTINAM (CNRS UMR 6213),\\ Universit\'e de Franche-Comt\'e,\\ 41 bis Avenue de l'Observatoire, BP1615, 25030 Besan\c con cedex, France.\\

J.-M. Vigoureux \\ Institut UTINAM (CNRS UMR 6213),\\ Universit\'e de Franche-Comt\'e, \\16 route de Gray, 25030 Besan\c con cedex, France.}

\end{center}
\maketitle \fontsize{12}{24}\selectfont

\begin{abstract}
In a recent paper \cite{vig2008} it has been suggested that the velocity of light and the expansion of the universe are two aspects of one single concept connecting space and time in the expanding universe. It has then be shown that solving Friedmann's equations with that interpretation (and keeping $c$ = constant)
can explain number of unnatural features of the standard cosmology (for example: the flatness problem, the problem of the observed uniformity in term of temperature and density of the cosmological background radiation, the small-scale inhomogeneity problem...) and leads to reconsider the Hubble diagram of distance moduli and redshifts as obtained from recent observations of type Ia supernovae without having to need an accelerating universe. 
In the present work we examine the problem of the cosmological constant. We show that our model can exactly generate $\Lambda$ (equation of state $P_\varphi = - \rho_\varphi c^2$ with $\Lambda \propto R^{-2}$) contrarily to the standard model which cannot generate it exactly. We also show how it can solve the so-called cosmic coincidence problem.
\end{abstract}

Keywords : {cosmology, cosmological constant, quintessence, accelerating universe, velocity of light, universal constant}\\
Pacsnumber : {98.80.Es, 98.80.Cq}

\section{Introduction}

\indent The constant $c$ was first introduced as the speed of light.
However, with the development of physics, it came to be understood
as playing a more fundamental role, its significance being not
directly that of a usual velocity (even though its dimensions are)
and one might thus think of $c$ as being a fundamental constant of
the universe. Moreover, the advent of Einsteinian relativity, the
fact that $c$ does appear in phenomena where there is neither light
nor any motion (for example in the fundamental equation $E$ = $m c^2$) and its double-interpretation in terms of "velocity of light" and of "velocity of gravitation" force everybody to associate $c$
with the theoretical description of space-time itself rather than
that of some of its specific contents. These considerations lead to connect $c$
to the geometry of the universe \cite{vig2008, vig2003}. In that interpretation, the velocity of light $c$ and the expansion of the universe are two aspects of one single concept connecting space and time in the expanding universe. This gives to $c$ a geometrical meaning and
makes of it a true {\em universal} quantity of the universe which can be defined from its size and its age without any other considerations.

Taking this into account, the fundamental equation for $c$ is
\begin{equation} \label{c}
c = \alpha \dot R 
\end{equation}

\noindent where $\alpha$ is a positive constant and where $R$ is the Robertson-Walker scale factor (a discussion concerning this equation can be found in \cite{vig2008}). It must be noted that here $c$, and consequently $\dot R $, are constant. With these assumptions, the theory of general relativity is unchanged, and consequently the Friedmann equations are still valid. 

In a previous paper \cite{vig2008} we have shown that solving Friedmann's equations with that interpretation of $c$ can explain number of unnatural features of the standard cosmology (the flatness problem, the problem of the observed uniformity in terms of temperature and density of the cosmological background radiation, the small-scale inhomogeneity problem...) and also leads to reconsider the Hubble diagram of distance moduli and redshifts as obtained from recent observations of type Ia supernovae \cite{Riess2004} \textit{without having to need an accelerating universe}.
In the present paper we are interested by other particular solutions of the usual Friedmann equations satisfying the additional eq(\ref{c}). 

In the $\Lambda$ Cold Dark Matter model ($\Lambda$CDM model), a cosmological constant $\Lambda$ is needed to explain the expansion acceleration \cite{Carroll2001}. A quintessence fluid is then often used to describe the source of the dark energy \cite{Carroll2001,Ratra1988,Wei1990,Benabed2001,Steinhardt1999}. Noting then $\rho_\varphi$ and $P_\varphi$ the mass density and the pressure of that fluid, it is then expected to get $P_\varphi = - \rho_\varphi c^2$ with $\rho_\varphi \propto R^{-n}$ with $n \not=0$ for the equation of state of the quintessence fluid. However, it has been shown that these two conditions cannot be simulaneously fulfilled in the standard model so that the standard model cannot generate perfectly the cosmological constant.

In the present model, \textit{where no acceleration is needed to explain the Hubble diagram of distance moduli and redshifts as obtained from recent observations of type Ia supernovae} \cite{vig2008}, a cosmological constant is nevertheless necessary to ensure the consistancy of the eq. (\ref{c}) with the state equation of the cosmic fluid. We consequently consider the problem and we show that our model can exactly generate the cosmological constant $\Lambda$. In fact, when using eq.(\ref{c}), we obtain \textit{as expected} (and contrarily to the $\Lambda$CDM model which cannot exactly find this solution), the equations $ P_{\varphi} = - \rho_{\varphi} c^2$ and $\rho_\varphi \propto R^{-2}$ for the equation of state of the quintessence fluid which generates $\Lambda$ and for its mass density respectively. The expression $\rho_\varphi \propto R^{-2}$ implies $\Lambda \propto R^{-2}$ and consequently means that the effective cosmological constant must vary with time. Such a result has been extensively discussed in the literature (see for example \cite{Ford1985,Ratra1988,Peebles1988}) and has been shown to be consistent with quantum cosmology \cite{Wei1990}). We also show how our model can solve the problem of the so-called "cosmic coincidence".

The structure of  this paper is as follows: Sec. 2 briefly recalls some important results of the standard cosmology that we need in what follows. Then we explain the problem of the cosmological constant in the standard model and in our. We finally show how eq.(\ref{c}) can solve the coincidence problem.

\section{The evolution of the cosmic fluid in the standard model}

	Our aim in this part is to recall some well-known results we need in what follows.
Let $\rho(t)$ and $P(t)$ be the mass density and the pressure of the cosmic fluid respectively. When there is no cosmological constant ($\Lambda = 0$), the two Friedmann's equations are

 \begin{equation} \label{fried1zero} 
 \left( \frac{\dot R}{R} \right)^2 = \frac{8 \pi G}{3} \rho - \frac{kc^2}{R^2} 
  \end{equation} 
  
  \begin{equation} 
  \frac{\ddot R}{R} = - \frac{4 \pi G}{3} \left( \rho + \frac{3 P}{c^2} \right)  \end{equation} 
  
\noindent where overdots are time derivatives. Noting as usually $P = \beta\rho c^2$ the above equation can also be written
  \begin{equation} \label{fried2zero} 
  \frac{\ddot R}{R} = - \frac{4 \pi G}{3} \rho(1 + 3\beta) \end{equation} 

\noindent Integrating these equations in the case of a flat universe ($k=0$), leads to the following results for $\rho$:

In a matter-dominated universe ($P=0$) one gets
   \begin{equation}  
   \rho \propto \frac{1}{R^3}.
   \end{equation}
   
In a radiation-dominated universe ($P = \frac{1}{3} \rho c^2$), one gets 
     \begin{equation}  
   \rho \propto \frac{1}{R^4}.
   \end{equation}
   
In the most general case ($P = \beta \rho c^2$ with $\beta >0$), one gets 
      \begin{equation}\label{ro}  
    \rho \propto \frac{1}{R^{3(\beta+1)}}.
   \end{equation}

 \section{The problem of the cosmological constant} 
\indent To explain the origin of the cosmological constant, some models introduce a quintessence fluid the mass density and the pressure of which (denoted $\rho_\varphi$ and $P_\varphi$) being thus to be included in the Friedmann's equations. Assuming then, as is usual, that the equation of state of the quintessence fluid has the form
\begin{equation} \label{varphi} 
  P_\varphi = \gamma \rho_\varphi c^2
   \end{equation} 
  (where the constant $\gamma <0$, which has to be determined, must be negative to get an anti-gravity), the two Friedmann's equations (\ref{fried1zero}) and (\ref{fried2zero}) become :

\begin{equation} \label{F1lambda} 
  \left( \frac{\dot R}{R} \right)^2 = \frac{8 \pi G}{3} (\rho + \rho_\varphi)  - \frac{kc^2}{R^2} 
 \end{equation} 
 
  \begin{equation} \label{F2lambda} 
 \frac{\ddot R}{R}  =  - \frac{4 \pi G}{3} \left(\rho (1 + 3\beta)+ \rho_\varphi(1 + 3\gamma)\right)
   \end{equation} 

\noindent Noting then
\begin{equation} \label{lambda11} 
\Lambda = 8 \pi G \rho_\varphi 
\end{equation}
in the first equation (eq.\ref{F1lambda}), and
\begin{equation} \label{lambdaprim} 
\Lambda = - 4 \pi G \rho_\varphi(1+3\gamma)
\end{equation}
\noindent in the second (eq.\ref{F2lambda}), one obtains the two usual Friedmann's equation with the cosmological constant $\Lambda$:

 \begin{equation} \label{fried1} 
 \left( \frac{\dot R}{R} \right)^2 = \frac{8 \pi G}{3} \rho - \frac{kc^2}{R^2} + \frac{\Lambda}{3}
  \end{equation} 
  
  \begin{equation} 
  \frac{\ddot R}{R} = - \frac{4 \pi G}{3} \left( \rho + \frac{3 P}{c^2} \right) + \frac{\Lambda}{3} \end{equation} 
  
\noindent Of course, $\Lambda$ must be the same in the two above equations, so that equations (\ref{F1lambda}) and (\ref{F2lambda}) are coherent (and consequently the quintessence fluid can generate the cosmological constant $\Lambda$) if and only if eqs(\ref{lambda11}) and (\ref{lambdaprim}) give the same value for $\Lambda$ that is to say if and only if
\begin{equation} \label{lambda12}
\gamma = -1
\end{equation}

\noindent or, by inserting this result inside eq.(\ref{varphi}), if and only if
\begin{equation} \label{eqstate}
P_{\varphi}= - \rho_{\varphi} c^2
\end{equation}

\noindent (At this point, it must be underlined that observations of supernovae show that $\gamma = -1.02^{+0.13}_{-0.19}$ \cite{Riess2004} as expected in eq.(\ref{lambda12})).

\section{The cosmological constant in the standard cosmology}
\indent Derivating eq.(\ref{F1lambda}) and inserting eq.(\ref{F2lambda}) into the result gives the equation
\begin{equation}  
 -3 \frac{\dot R}{R} \left(\rho(1+\beta) + \rho_\varphi (1+\gamma) \right) = \dot \rho + \dot \rho_\varphi
 \end{equation} 

\noindent the solution of which for $\rho_\varphi$ is
\begin{equation}  
\rho_\varphi \propto \frac{1}{R^{3(\gamma+1)}}
\end{equation}  

The coherence condition $\gamma = -1$ (eq.\ref{lambda12}) then leads to $ \rho_\varphi = Cst $, and, using eq.(\ref{lambda11}), to $\Lambda = Cst$. This result ($ \rho_\varphi = Cst $) makes $\Lambda$ to be a pure constant but in that case the quintessence fluid doesn't dilute ($ \rho_\varphi = Cst $) when the universe expands which is rather suprising for a fluid which would, on the contrary, verify $\rho_\varphi \propto R^{-n}$ with $n \not=0$. The fact that the two conditions $\gamma = -1$ and $\rho_\varphi \propto R^{-n}\not=Cst$ cannot be simultaneously fulfilled shows that the standard model cannot generate perfectly the cosmological constant in the Friedmann's equations.\\

\section{The cosmological constant in the present model}

\indent Our aim now is to show that, contrarily to what is found with the standard model, the two conditions $P_{\varphi}= - \rho_{\varphi} c^2$ ($\gamma = -1$) and $\rho_\varphi \propto R^{-n}$ with $n \not=0$ can be simultaneously fulfilled when using eq.(\ref{c})

\indent When using eq.(\ref{c}), the two Friedmann's equations (\ref{F1lambda}) and (\ref{F2lambda}) become:
 \begin{equation} \label{fried1v} 
\left( \frac{c}{\alpha R} \right)^2 = \frac{8 \pi G }{3} \frac{\rho + \rho_\varphi}{1 + k \alpha^2} 
 \end{equation} 
 
\noindent and
\begin{equation} \label{lambdabeta}
 0 = \rho(1+3 \beta)+ \rho_\varphi (1+3 \gamma)
\end{equation} 

\noindent Calculating $\rho_\varphi$ from the second equation and inserting the result with $\gamma = -1$ into eq.(\ref{fried1v}) give

\begin{equation}
\label{rho}
\rho = \frac{c^2}{4 \pi G} \frac{(1 + k\alpha^2)}{\alpha^2} \frac{1}{1+\beta} \frac{1}{R^2} 
\end{equation} 

\begin{equation}\label{roro}
 2\rho_\varphi  = \rho(1+ 3\beta)
\end{equation}

\noindent Using then eq.(\ref{lambda11}) we find

\begin{equation}
\label{rho}
\rho = \frac{c^2}{4 \pi G} \frac{(1 + k\alpha^2)}{\alpha^2} \frac{1}{1+\beta} \frac{1}{R^2} 
\end{equation}\begin{equation} 
\Lambda = \frac{c^2 (1 + k\alpha^2)}{\alpha^2} \frac{1+ 3\beta}{1+\beta} \frac{1}{R^2} 
\end{equation}

\noindent We thus find
\begin{equation} \label{rhoR2}
\rho \propto \frac{1}{R^2} 
\end{equation}
\noindent and, using (\ref{roro})
\begin{equation} \label{rhoprimR2}
\rho_\varphi\propto \frac{1}{R^2} 
\end{equation}
\noindent and
\begin{equation} \label{lambdaR2}
\Lambda \propto \frac{1}{R^2} 
\end{equation}
{\em whatever may be $\beta \neq -1$}, that is to say, whatever may be the equation of state of the cosmic fluid, whereas in the standard cosmology $\rho$ does depend on $\beta$ ($\rho \propto \frac{1}{R^{3(\beta+1)}}$). Contrarily to what is obtained in the standard cosmology, the present model thus do fulfil the two conditions $\gamma = -1$ and $\rho_\varphi \propto {R^{-n}}$ (with $ n \not=0$, here with $n = 2$) simultaneously. It can consequently explain the origin of the cosmological constant with a quintessence fluid which dilutes, as expected, when the universe expands.

That result (\ref{lambdaR2}) of course means that $\Lambda$ varies with time. Such a cosmology with a time variable cosmological constant has been extensively discussed in the litterature \cite{Ford1985,Ratra1988} (Our result (\ref{lambdaR2}) joins that of ref.\cite{Ozer1987} in a different context). Moreover, it must be underlined here that eq.(\ref{lambdaR2}) has been shown to be in conformity with quantum cosmology \cite{Wei1990} and that a varying cosmological constant leads to no conflict with existing observations \cite{Riess2004}.

Let us add that a key problem with the quintessence proposal is to explain the so-called cosmic coincidence problem. In fact, it appears that the conditions in the early universe have to be set very carefully in order the densities $\rho$ and $\rho_\varphi$ are comparable today. This problem is clarified in the present approach. In fact, eq.(\ref{roro}) shows that, as expected, $\rho$ and $\rho_\varphi$ have the same order of magnitude \textit{at all times}. Moreover, it also shows that the two energy densities $\rho$ and $\rho_\varphi$ are  exactly equal when $\beta = \frac{1}{3}$ that is to say in a radiation dominated epoch. 

We can also note that eq.(\ref{c}) can also explain why the mass density of the cosmic fluid is so near the critical density $\rho_{c}$: noting in fact, that the critical density is given by
 \begin{equation} \label{rocritique} 
 \left( \frac{\dot R}{R} \right)^2 = \frac{8 \pi G}{3} \rho_{c} 
  \end{equation} 

\noindent and that eq.(\ref{fried1zero}) \textit{with eq.(\ref{c})} gives in the case of a spherical universe (k = 1),
\begin{equation} \label{univspher} 
 \left( \frac{\dot R}{R} \right)^2 (1 + \alpha^{2}) = \frac{8 \pi G}{3} \rho 
  \end{equation} 

 \noindent we easily find that the present model gives
 \begin{equation} \label{rosurroc} 
\rho = \rho_{c} (1 + \alpha^{2}) 
  \end{equation} 

\noindent At this point, it can be usefull to recall (see \cite{vig2008}) that \textit{in the present model a spherical universe (k = 1) must appear to be flat but with a smaller mass than expected} in the standard cosmology.

\section{Conclusions}
To conclude, our model
\begin{itemize} 
        \item leads, as expected, to the equation of state $P_\varphi = \gamma \rho_\varphi c^2 = - \rho_\varphi c^2$ for the fluid which generates the cosmological constant.
        \item it agrees with observations of supernovae which show that $\gamma = -1.02^{+0.13}_{-0.19}$ \cite{Riess2004},
          \item contrarily to the standard cosmology, it agrees with the constraint needed for the sake of coherence when introducing a cosmological constant in the two Frieddmann's equations (we mean it can perfectly generate the cosmological constant). 
        \item it also conducts to the equation $\rho_\varphi \propto R^{-n}$ with $n\neq 0$ (here $n=2$) as expected for a fluid which of course must dilute when the universe expands. 
        \end{itemize}
\noindent Two appealing features of our model are that it can simultaneously accommodate the equation of state $P_\varphi = - \rho_\varphi c^2$ of the quintessence fluid which generates $\Lambda$ with a varying density $\rho_\varphi \propto \frac{1}{R^2}$. That last result has been shown to be in conformity with quantum cosmology \cite{Wei1990} and a $\frac{1}{R^2}$ varying cosmological constant also appears to lead to no conflict with existing observations \cite{Riess2004}. As explained above, the present model also explains why $\rho$ and $\rho_\varphi$ are comparable and thus solve the coincidence problem. Using (\ref{roro}) it also shows why the present value of the critical energy density $\rho_c$ is so near the present energy density $\rho$ of the universe (let us recall that, as explained in \cite{vig2008}, when using eq.(\ref{c}) the condition $\rho = \rho_c$ doesn't need to be in a flat universe).

Noting that eq.(\ref{c}) leads to other interesting ways to explain unnatural features of the standard cosmology  \cite{vig2008} (for example : the flatness problem, the problem of the observed uniformity in term of temperature and density of the cosmological background
radiation, the small-scale inhomogeneity problem) and that it also permits to reconsider the Hubble diagram of distance moduli and redshifts as obtained from recent observations of type Ia supernovae without having to need an accelerating universe, we may think that the present results can lead to a satisfactory cosmology.

\end{document}